\documentclass{sigchi}

\toappear{
\textcopyright2019. Copyright for the individual papers remains with the authors. \\
Copying permitted for private and academic purposes. \\
MILC `19, March 20, 2019, Los Angeles, CA, USA \\
}

\usepackage{balance}       % to better equalize the last page
\usepackage{graphics}      % for EPS, load graphicx instead 
\usepackage[T1]{fontenc}   % for umlauts and other diaeresis
\usepackage{txfonts}
\usepackage{mathptmx}
\usepackage[pdflang={en-US},pdftex]{hyperref}
\usepackage{color}
\usepackage{cite}
\usepackage{booktabs}
\usepackage{textcomp}
\usepackage{multirow}
\usepackage{hhline}

\usepackage{microtype}        % Improved Tracking and Kerning
\usepackage{ccicons}          % Cite your images correctly!
\usepackage{todonotes}

\def\plaintitle{A Minimal Template for Interactive Web-based Demonstrations of Musical Machine Learning}

\def\emptyauthor{}
\def\plainkeywords{Musical interface; web; latent space; deep learning}

\makeatletter
\def\url@leostyle{%
  \@ifundefined{selectfont}{
    \def\UrlFont{\sf}
  }{
    \def\UrlFont{\small\bf\ttfamily}
  }}
\makeatother
\def\pprw{8.5in}
\def\pprh{11in}

\definecolor{linkColor}{RGB}{6,125,233}
\definecolor{todoColor}{RGB}{255,0,0}
\hypersetup{%
  pdftitle={\plaintitle},
  pdfauthor={\emptyauthor},
  pdfkeywords={\plainkeywords},
  pdfdisplaydoctitle=true, % For Accessibility
  bookmarksnumbered,
  pdfstartview={FitH},
  colorlinks,
  citecolor=black,
  filecolor=black,
  linkcolor=black,
  urlcolor=linkColor,
  breaklinks=true,
  hypertexnames=false
}

\begin{document}

\title{\plaintitle}
\numberofauthors{1}
\author{%
  \alignauthor{Vibert Thio, Hao-Min Liu, Yin-Cheng Yeh, and Yi-Hsuan Yang\\
    \affaddr{Research Center for Information Technology Innovation, Academia Sinica, Taipei, Taiwan}\\
    \email{\{vibertthio,~paul115236,~ycyeh,~yang\}@citi.sinica.edu.tw}}\\
}

\maketitle

\begin{abstract}
New machine learning algorithms are being developed to solve problems in different areas, including music. Intuitive, accessible, and understandable demonstration of the newly built models could help attract the attention of people from different disciplines and evoke discussions.
%, and accordingly foster an active research community. 
However, we notice that it has not been a common practice for researchers working on musical machine learning to demonstrate their models in an interactive way. To address this issue, we present in this paper an template that is specifically designed to demonstrate symbolic musical machine learning models on the web. The template comes with a small codebase, is open source, and is meant to be easy to use by any practitioners to implement their own demonstrations. Moreover, its modular design facilitates the reuse of the musical components and accelerates the implementation. We use the template to build interactive demonstrations of four exemplary music generation models. We show that the built-in interactivity and real-time audio rendering of the browser make the demonstration easier to understand and to play with. It also helps researchers to gain insights into different models and to A/B test them.
\end{abstract}



\section{Introduction}
Recent years have witnessed great progress in applying machine learning (ML) to music related problems, such as 
%source separation~\cite{liu18icmla}, %Jansson2017
thumbnailing~\cite{huang18tismir}, 
music generation~\cite{lpd,Magenta:MusicVAE,GAN:DRUM:UI}, and 
style transfer~\cite{lu19aaai}. 
%However, to date it is still rare to see a demonstration for a musical machine learning model. 
To demonstrate the result of such musical machine learning models, researchers usually 
%When it comes to a musical presentation, it is a lot easier to 
put the audio output as the result on the accompanying project websites. This method works well in the early days. However, as the ML models themselves are getting more complicated, some concepts of the algorithms may not be clearly expressed with only static sounds.

In the neighboring field of computer vision, many interactive demonstrations of ML models have been developed recently. Famous examples include DeepDream~\cite{google:deep_dream}, image/video style transfer~\cite{style_transfer,RDB18}, and DCGAN~\cite{DCGAN}. 
%there are a bunch of interactive applications built around those models. 
These interactive demos provoke active discussions and positive anticipation about the  technology.
Nevertheless, the demonstration of musical machine learning models is not as easy in the case of computer vision, due to the fact that it involves audio rendering (i.e., we cannot simply use images for demonstration). Web Audio API, a high-level JavaScript API for processing and synthesizing audio in web applications, was published only in 2011, which is not far from now compared to WebGL and other features of the browser. Furthermore, interactivity is needed to improve understandability and create engaging experiences. 

Musical machine learning is gaining increasing attention. We believe that if more people from other fields, such as art and music, start to appreciate the new models of musical machine learning, it is easier to create an active community and to stimulate new ideas to improve the technology.

The goal of this paper is to fulfill this need by building and sharing with the community a template that is designed to demonstrate ML models for symbolic-domain music processing and generation, in an interactive way. Therefore, The template is also open-source on GitHub (\url{https://github.com/vibertthio/musical-ml-web-demo-minimal-template}).

%The goal of this paper is to fulfill this need by building and sharing a template that helps building interactive demonstrations for ML models of symbolic-domain music with the community.

\section{Related Works}

\subsection{Audio Rendering in Python}

When it comes to testing or interacting with the musical machine learning models, the output of the models must be rendered as audio files or streams to be listened to by humans. Most researchers in the field nowadays use Python as the programming language for model implementation because of the powerful ML and statistical packages built around it.
For example, \texttt{librosa}~\cite{librosa} is a Python package often used for audio and signal processing. 
%It is often used in the field of music information retrieval (MIR), which is strongly related to music machine learning. 
It includes functions for spectral analysis, display, tempo detection, structural analysis, and output. Many interactive demonstrations %and documentation 
are built with librosa on the Jupyter Notebook. However, a major drawback of this approach is that the audio files have to be sent over the Internet for demonstration, which can be slow sometimes depending on the network connection bandwidth.
%is slow and not efficient, compared to symbolic representation, like Musical Instrument Digital Interface (MIDI) data. Despite the wide range of its functionality, librosa doesn`t support the symbolic output.

Another widely-used Python package is  \texttt{pretty\_midi}~\cite{pretty_midi}, which is designed for manipulation of symbolic-domain data such as Musical Instrument Digital Interface (MIDI) data.
It could be used as a tool to render the symbolic output of a musical machine learning model, such as a melody generation model~\cite{performanceRNN}.
The problem is that after getting the result as a MIDI file, the user still has to put it into a digital audio workstation (DAW) to synthesize the audio waveform from the MIDI. For better listening experience, the researcher still has to synthesize the audio files offline and then send the audio files over the Internet for demonstration.

%There are some existing tools, but most of them are either monolithic or hard to customize. 

Different from prior works, we propose to use Tone.js~\cite{tone}, a JavaScript framework for rendering MIDI files into audio directly in the browser on the client side. This turns out to be a much more efficient way to demonstrate a symbolic musical ML model. It also helps build an interactive demo.

\subsection{Interactive Musical Machine Learning}

Similar to our work, Vogl \emph{et al.}~\cite{GAN:DRUM:UI} introduced an interactive App for drum pattern generation based on ML. They used generative adversarial networks (GAN) \cite{GAN} as the generative model, which is trained on a MIDI dataset. The user interface consists of some classic sequencer with an x/y pad that controls the complexity and the loudness of the drum pattern generated. Additionally, controls for genre and swing are said to be provided. However, both the demo and its source code cannot be found online currently.
It is not clear whether the App is built on iOS, Android, or the Web.

Closely related to our project is the MusicVAE model~\cite{Magenta:MusicVAE} presented by Magenta, Google Brain Team. MusicVAE is a generative recurrent variational autoencoder (VAE) model that can generate melodies and drum beats. Importantly, the authors also released a JavaScript package called \texttt{Magenta.js} (\url{https://github.com/tensorflow/magenta-js/})~\cite{Magenta:js}, to make their models more accessible. They also provide some pre-trained models of MusicVAE along with other ones. There are several interactive demos using the package, as can be found on their website~\cite{Magenta:demos}.
%and the models were featured  
%including `Melody Mixer,' `Beat Blender,' and `Latent Loops'~\cite{Magenta:demos}.
Most of them are well designed, user-friendly, and extremely helpful for understanding the models. Yet, the major drawback is that the codebase of the project is a \emph{monolithic} one  \cite{47040} and is therefore quite big.\footnote{A monolithic repo is defined in \cite{47040} as: ``a model of source code organization where engineers have broad access to source code, a shared set of tooling, and a single set of common dependencies. This standardization and level of access are enabled by having a single, shared repo that stores the source code for all the projects in an organization.'' This is the case of the Google Magenta project.} Users may not easily modify the code for customization. For example, because Magenta uses Tensorflow as the backbone deep learning framework, it is hard for PyTorch users to use \texttt{Magenta.js}.

%switch the models in the package, and the codebase of the project is monolithic. Therefore, modifying the source code for customization is comparably difficult.

\section{Template Design}

In this paper, we present a minimal template, which is simple, flexible, and designed for interactive demonstration of symbolic musical machine learning models on the web.

We choose the web as the platform for the demonstrations for several reasons. First, it is convenient as the user only has to open a browser or click the hyperlink to play with the models. Second, it is inherently interactive. The system can utilize a plenty of forms of interaction available in the browser to create the specific user experience.

\subsection{Requirements}

In the design process, we have prioritized some crucial qualities. First, we made the structure of the design as simple as possible. 
In most cases, the demo is for a proof-of-concept rather than to showcase a ready-to-sell product. Hence, we desire that a person with basic knowledge of Python and JavaScript could understand our template within a short period of time so that the template can serve as a minimal starting point.
%it in less than a few hours.

Second, the codebase should be small, so that transplanting a new model into this template is easier. Moreover, a small codebase also makes it easier to debug.
%When testing a model in the development phase, it is essential that the developer can understand every detail of the process. Otherwise, the source of bugs might be hard to identify. Moreover, when the codebase is small enough, 

Third, the audio rendering must be interactive and real-time. The demonstrations must be responsive to some inputs from the user so that the user could understand the model by knowing how it works in several different ways. As for researchers, if the result could be rendered instantly, it would be easier to A/B test different designs of models or parameters.

Finally, we want the components of the template to be modular, so that they can be reused and recombined easily. Such components may include, e.g.,  \texttt{chord progression}, \texttt{pianoroll}, \texttt{drum pattern}, and \texttt{sliders}.
%, to name a few. The 
Practitioners can build their own demonstrations based on these components.

\begin{figure}
\centering
  \includegraphics[width=0.9\columnwidth]{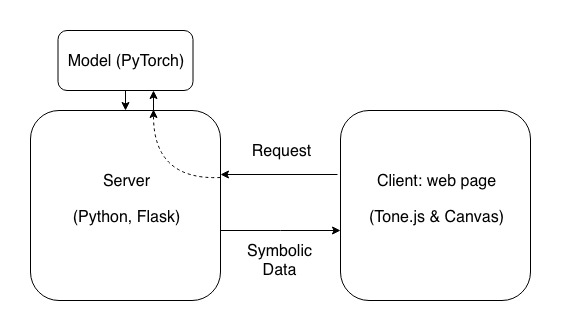}
  \caption{The schematic diagram of the proposed template.}~\label{fig:figure1}
\end{figure}

\begin{figure*}
\centering
  \includegraphics[width=2.0\columnwidth]{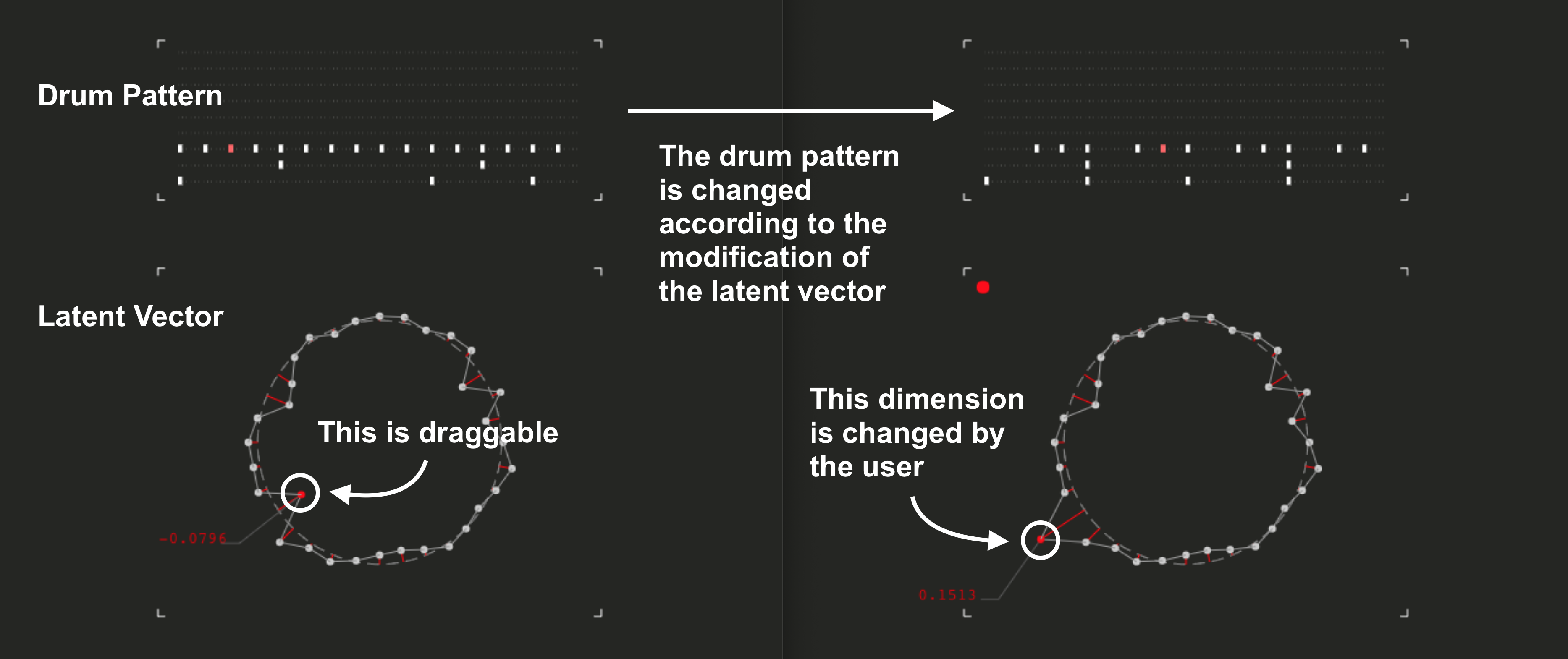}
  \caption{The Latent Inspector. Left: a snapshot of the demonstration. The upper half is an editable drum pattern display, whereas the bottom half is a graph showing the $N$-dimensional latent vector (here $N=32$) of the VAE model. Each vertex of the graph can be adjusted independently, to change the latent vector and accordingly the drum pattern.
  This is exemplified by the snapshot shown on the right.
  %, where an entry close to the bottom left is adjusted. 
  Although not shown in the figure, we can also click on the drum pattern display to modify the drum pattern directly, which would also change the latent vector accordingly.}~\label{fig:drumvae}
\end{figure*}

\subsubsection{System Architecture}

As shown in  Figure~\ref{fig:figure1}, the system consists of three parts: a musical machine learning model, a server, and a client. When a user opens the URL of the demonstration site, it will load the client program into the browser and render the basic interface. The client program will send a request to the server to fetch the data. The server program will parse the request and use the function based on the model to make the corresponding output and send it back to the client to render the audio effect.

\subsection{Server}
We used \texttt{Flask} (\url{http://flask.pocoo.org/})~\cite{FLASK}, a lightweight web application framework, to build the server. % to deploy the model.
Flask only handles essential core functions to build a web server, such as representational state transfer (REST) handling the request from the client. Therefore, we can build the server without any redundant elements but focus on the function of the model. As a result, the server template code has only about 150 lines, excluding the model implementation part.

\subsection{Client}
Several technologies have been built to render the real-time audio output since the Web Audio API was released in 2010~\cite{Web_Audio_API}. \texttt{Tone.js} (\url{https://tonejs.github.io/})~\cite{tone} is such a framework for creating interactive music in the browser. It provides simple workflows for synthesizing audio from oscillators, organizing the audio samples, musician-friendly API, and timeline schedule. It makes the development of real-time rendering from the output of the ML model much easier. Added in the new standard HTML5, the HTML canvas element can be used to draw dynamic graphics via writing JavaScript~\cite{canvas}. Thus, we use JavaScript canvas with \texttt{Tone.js} to create audio and visual experience coherently.

\begin{figure*}
\centering
  \includegraphics[width=2.0\columnwidth]{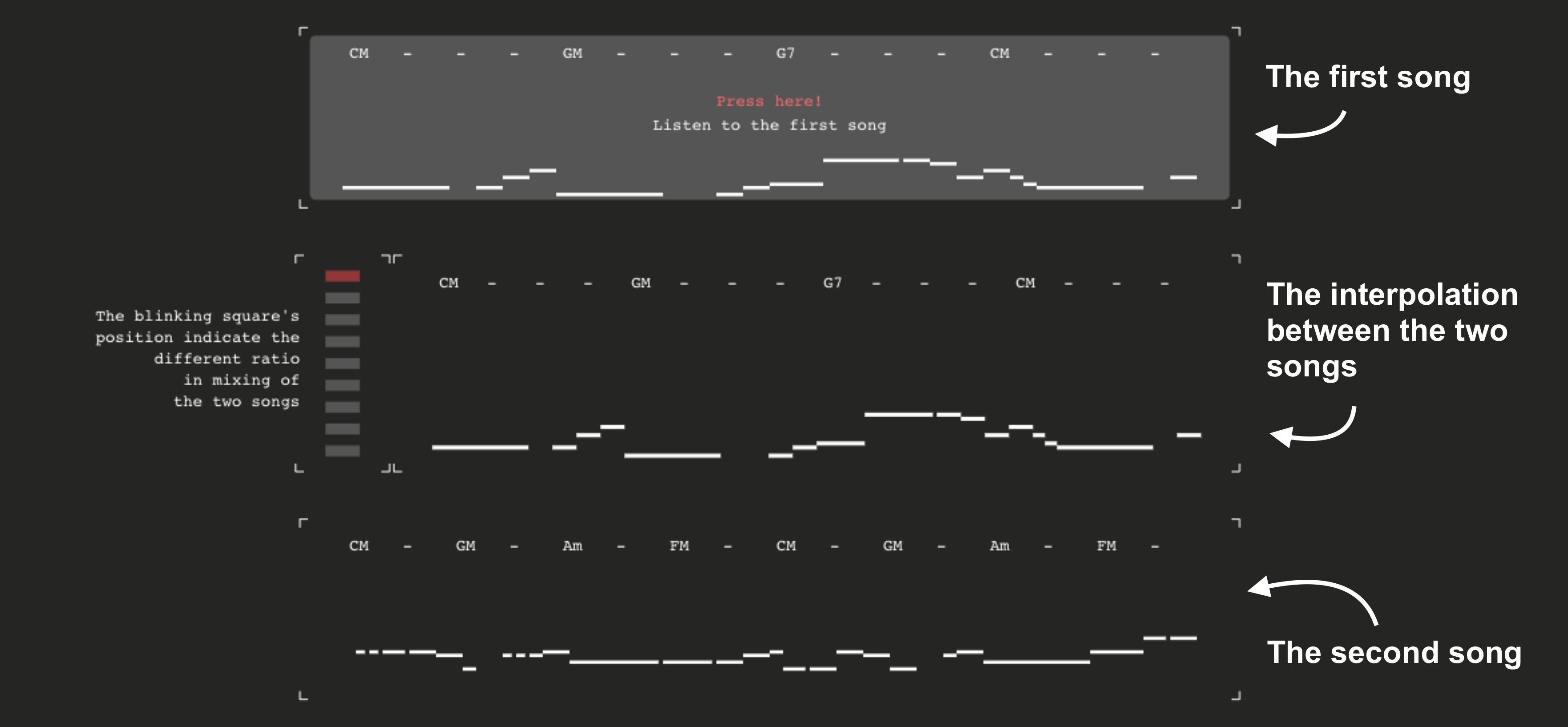}
  \caption{The Song Mixer. The top and bottom panels display the melody and the chords of the first and the second song.  In the middle is the interpolation between the two songs. 
  We add visual aids to guide the user to interact with the App. For example, the top panel is highlighted in this figure to invite the user to listen to the first song, before the second song and then the interpolations.}~\label{fig:songs}
\end{figure*}

The modularization is taken care of in the design of the interface 
%(see Figures~\ref{fig:drumvae}--\ref{fig:m2c}).
(see Figures~\ref{fig:drumvae}--\ref{fig:tuningturing}).
%(see Figures~\ref{fig:drumvae}, ~\ref{fig:songs}, and ~\ref{fig:m2c}).
The layout of the user interfaces is implemented as a grid system. This speeds up the design process because it simplifies the choices for the positions of the elements and the margins between them. The recurring elements, such as \texttt{pianoroll} and \texttt{drum pattern} display, are implemented based on object-oriented principles, thus they can be re-used easily.
%in different projects.
See Table \ref{tab:table1} for a summary.

\section{Demonstrations Design}

We built four different demonstrations based on the proposed template. We call them `Latent Inspector,' `Song Mixer,' `Comp It,' and `Tuning Turing.' Each of them was designed to serve one exact purpose and demonstrate a single idea based on the musical machine learning model. The classes of the models are not limited to certain ones. For instance, the first two demonstrate the musical machine learning models based on VAE~\cite{vae}. In contrast, the last two are mainly based on a recurrent neural network (RNN). Also, the type of instruments could be different. For example, the first one is about percussion and the other three are about melody. This is designed deliberately to show the general purpose of this template. We aim to make them more understandable and interesting by adding interactivity, interface design, and visual effects.

\subsection{Latent Inspector with DrumVAE}
DrumVAE is an original work. It uses VAE for generating one-bar drum patterns. Drum patterns are represented using the \texttt{pianoroll} format \cite{pypianoroll} with 96 time-step per bar. It compresses (or encodes) the drum patterns into a latent space via a bidirectional gated recurrent unit (BGRU) neural network~\cite{DBLP:journals/corr/ChoMGBSB14}. The outputs from BGRUs are used as mean and variance of a Gaussian distribution. A latent vector is sampled from the Gaussian distribution. We apply the similar but reverse structure of the encoder in the decoder and pass the latent vector into it to reconstruct the drum patterns. It is trained on one-bar drum patterns collected from the Lakh Pianoroll Dataset~\cite{lpd}, considering the following nine drums: kick drum, snare drum, closed/open hi-hat, low/mid/high toms, crash cymbal, and ride cymbal.

The Latent Inspector, shown in  Figure~\ref{fig:drumvae}, lets the user modify the latent vector of a drum pattern displayed in the browser to find out how the drum pattern will alter correspondingly. On the other hand, the user can also modify the drum pattern to observe the changes in the latent vectors.

X/Y pads are used in other works to explore the latent space. Yet, the dimension of the latent vectors used in practice is usually larger than two.
% As a result, an X/Y pad cannot explore all the dimensions at the same time. An X/Y pad also lacks visual expressiveness. 
%but also difficult to understand.
As a result, we designed a circular diagram which can represent high dimensional data. As shown in Figure~\ref{fig:drumvae}, the latent vector or our DrumVAE model has dimension $N=32$. Since the effect of every dimension should be symmetrical in the latent vector of DrumVAE, using circular diagram can eliminate the terminal point of the line chart.

It is possible to further improve the UI by adding conditional functionalities, to give each vertex some musical or semantic meaning.
While this can be a future direction, we argue that the current is also interesting---
%Although some may argue that the latent space without conditional training is not musically meaningful, Latent Inspector is designed as a helping tool to observe the behavior of the model. As 
for musicians, it is sometimes more interesting to have a bunch of knob of unknown functionalities to play with.
%, rather modifications than some well-designed parameters.

The demo website of  Latent Inspector can be found at \url{http://vibertthio.com/drum-vae-client/public/}.

\subsection{Song Mixer with LeadSheetVAE}

LeadSheetVAE is another model we recently developed~\cite{liu18ismir-lbd}.
It is also based on a VAE, but it is designed to deal with lead sheets instead of drum patterns.
A \emph{lead sheet} is composed of a melody line and a sequence of chord labels~\cite{liu18ismir-lbd}. We consider four-bar lead sheets here.  
Melody lines and chord sequences are represented using one-hot vectors and chroma vectors, respectively. It resembles the structure of DrumVAE, but the main difference is that by the end of the encoder the output of the two BGRUs (one for melody and one for chords) are concatenated and passed through few dense layers for calculating the mean and variance for the Gaussian distribution. In the decoder, we apply two unidirectional GRUs to reconstruct the melody lines and chord sequences. 
The model is trained on the TheoryTab dataset~\cite{liu18ismir-lbd} with 31,229 four-bar segments of lead sheets featuring different genres.
%After training this model on TheoryTab dataset, we could pick two existing songs or create one from scratch and explore the interpolation by sampling out latent vectors in between from the embedding space.
LeadSheetVAE can generate new lead sheets from scratch, but we use it for generating interpolations here.

The Song Mixer, shown in Figure~\ref{fig:songs}, takes two existing lead sheets as input and shows the interpolations of them generated by LeadSheetVAE. Similarly, a user can modify the melody or chords using the upper panel, or choose other lead sheets from our dataset,
to see how it affects the interpolation.

The aim of this demo is to make the interpolation understandable. Therefore, we build interactive guidance with visual cues through the process to make sure the user grasp the idea of lead sheet interpolation.
The demo website of Song Mixer can be found at \url{http://vibertthio.com/leadsheet-vae-client/}.

Evaluating the quality of interpolations generated by general VAE models (not limited to music-related ones), and many other generative models, has been known to be difficult. A core reason is that there is no ground truth for such interpolations. Song Mixer makes it easy to assess the result of musical interpolations. Moreover, with the proposed template, it is easy to extend Song Mixer to show the interpolation produced by two different models side-by-side and in-sync in the middle of the UI. This facilitates A/B testing the two different models with a user study.

\begin{figure*}
\centering
  \includegraphics[width=2.0\columnwidth]{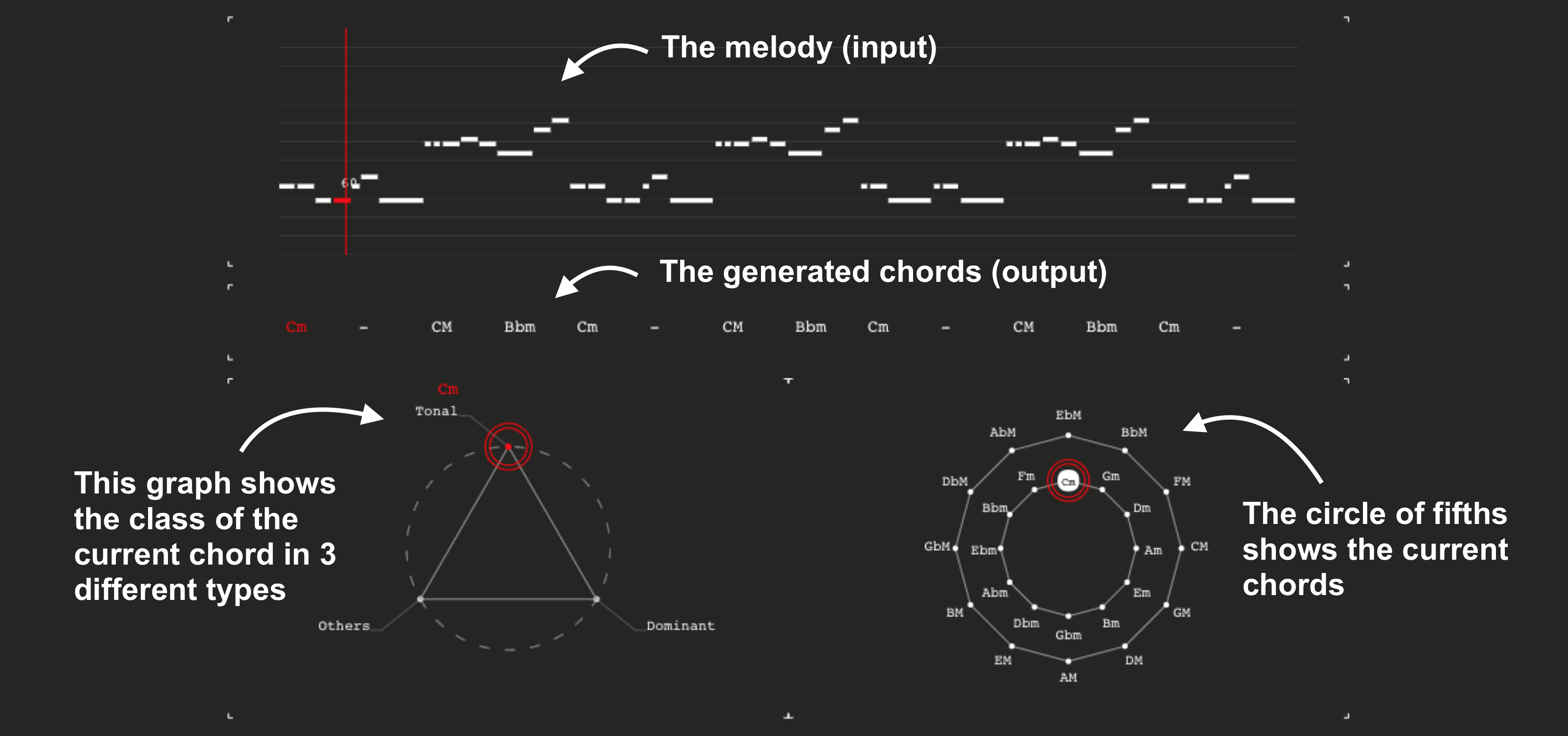}
  \caption{Comp It. The upper half is the editable melody and the chords predicted by the underlying melody harmonization model. In the lower left is a graph showing which class the current chord belongs to. In the lower right is a graph showing the position of the current chord on the so-called circle of fifths. The current melody note and chord being played are marked in red. }~\label{fig:m2c}
\end{figure*}

\subsection{Comp It \& Tuning Turing with MTRNNHarmonizer}
Finally, MTRNNHarmonizer is another new model that we recently developed.\footnote{More details of the model will be provided in a forthcoming paper.}
It is an RNN-based model for adding chords to harmonize a given melody. In other words, given a melody line, the model produces a chord sequence to make it a lead sheet.
The model is special in that it takes a multi-task learning framework to predict not only the chord label but also the chord's functional harmony, for a given segment of melody (half-bar in our implementation). 
Taking the functional harmony into account makes the model less sensitive to the imbalance of different chords in the training data. 
Furthermore, the chord progression can have the phrasing that better matches the given melody line. 

%\textcolor{red}{We use MTRNNHarmonizer to build two demonstrations.} 
Similar to the two aforementioned demos, Comp It allows a user to modify the melodies displayed in the browser to find out how this will alter the chord progression correspondingly. Furthermore, 
as shown in Figure~\ref{fig:m2c}, we add a triangular graph and an animated \emph{circle of fifths}~\cite{circleoffifth} graph to visualize the changing between different chord classes.
The triangular graph displays the chord class of the chord being played, covering tonal, dominant, and sub-dominant.
The circle of fifths graph, on the other hand, organizes the chords in a way that reflects the ``harmonic distance'' between chords~\cite{bello05ismir}.
These two graphs make it easier to study the chord progression generated by the melody harmonization model, which is MTRNNHarmonizer here but can be other models in other implementations.  

\begin{figure}
\centering
  \includegraphics[width=0.9\columnwidth]{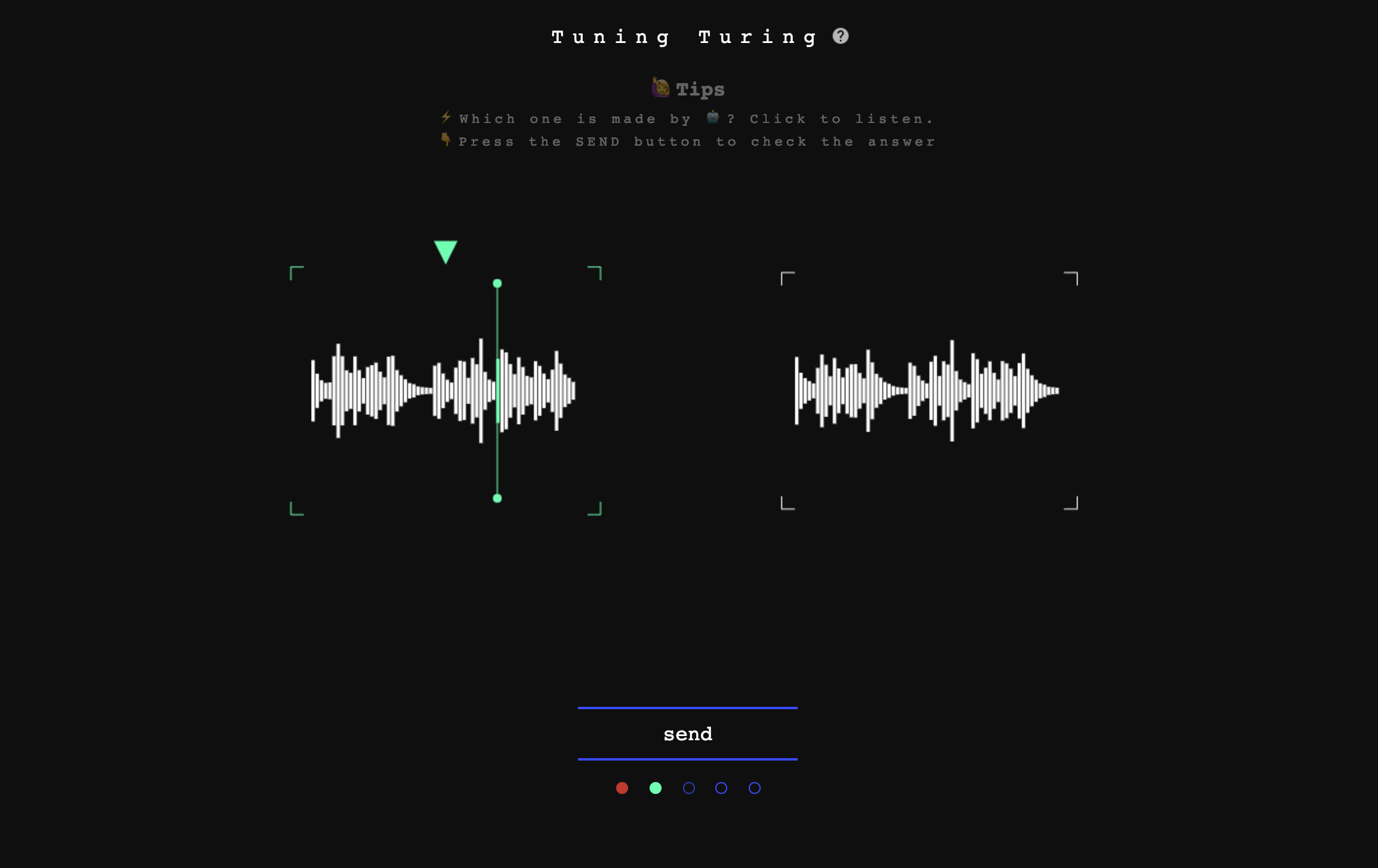}
  \caption{Tuning Turing. Two different kinds of harmonization for a single melody are rendered on the page. The player has to pick out the one generated by the algorithm and send the result after choosing with the mouse.}
  ~\label{fig:tuningturing}
\end{figure}

Furthermore, we made a simple \emph{Turing game} for the model, called ``Tuning Turing.'' 
As shown in Figures~\ref{fig:tuningturing},
the player has to pick out the harmonization generated by the model from two music clips. There are both ``practice mode'' and ``challenge mode.'' The former has 6 fixed levels. In the ``challenge mode,'' the player can keep playing until three wrong answers.

The demo website of Comp It and Tuning Turing can be found at \url{http://vibertthio.com/m2c-client/} and \url{http://vibertthio.com/tuning-turing/} respectively.

\section{Availability}
Supplementary resources including open source code will be available at the GitHub repos (\url{https://github.com/vibertthio}), including the template (\url{https://github.com/vibertthio/musical-ml-web-demo-minimal-template}), the interfaces, and the ML models.

\section{Conclusion}
% and Future Work

This paper presents an open-source template for creating an interactive demonstration of musical machine learning on the web along with four exemplary demonstrations. The architecture of the template is meant to be simple and the codebase is small so that other practitioners can implement their models with it within a short time. The modular design makes the musical component reusable. The interactivity and real-time audio rendering of the browser make the demonstration easier to understand and to play with. However, we try to elaborate the quantitative aspects of the project without quantitative analysis. For future work, we will run user studies to validate the effectiveness of these projects.
%Musical machine learning is growing fast. While great progress is being made in recent years by computer science researchers, they are not well recognized by other communities, for example, artists and musicians. 
With more intuitive, accessible, and understandable demonstrations of the new models, we hope new people might be brought together to form a larger community to stimulate new ideas.

\begin{table}
\centering
\begin{tabular}{ |l|p{5cm}| }
    \hline
    \textbf{Demonstration} & \textbf{Modules} \\
    \hline
    \multirow{4}{*}{Latent Inspector}
    & audio rendering (sample) \\ \cline{2-2}
    & editable pianoroll (drum) \\ \cline{2-2}
    & editable latent vector (circular) \\ \cline{2-2}
    & **radio panel (genre selection) \\ \cline{2-2}
    \hline
    \multirow{4}{*}{Song Mixer}
    & audio rendering (synthesize) \\ \cline{2-2}
    & editable pianoroll (melody) $\times$ 3 \\ \cline{2-2}
    & chord visualization (text) \\ \cline{2-2}
    & radio panel (interpolations selection) \\ \cline{2-2}
    \hline
    \multirow{3}{*}{Comp It}
    & audio rendering (synthesize) \\ \cline{2-2}
    & editable pianoroll (melody) \\ \cline{2-2}
    & chord visualization (text, function, circle of fifths) \\ 
    \hline
    \multirow{3}{*}{Tuning Turing}
    &  audio rendering (sample)\\
    \cline{2-2}
    &  waveform visualization\\ 
    \hline
\end{tabular}
\caption{Some of the modules are reused by more than one demonstration. For example, three of them uses ``editable pianoroll''. In our implementation, We reuse the modules to reduce the development effort. Therefore, it is useful and convenient to develop new demos with these modules. ** indicates that a  feature that has not been implemented yet.}~\label{tab:table1}
\end{table}

% Balancing columns in a ref list is a bit of a pain because you
% either use a hack like flushend or balance, or manually insert
% a column break.  http://www.tex.ac.uk/cgi-bin/texfaq2html?label=balance
% multicols doesn't work because we're already in two-column mode,
% and flushend isn't awesome, so I choose balance.  See this
% for more info: http://cs.brown.edu/system/software/latex/doc/balance.pdf
%
% Note that in a perfect world balance wants to be in the first
% column of the last page.
%
% If balance doesn't work for you, you can remove that and
% hard-code a column break into the bbl file right before you
% submit:
%
% http://stackoverflow.com/questions/2149854/how-to-manually-equalize-columns-
% in-an-ieee-paper-if-using-bibtex
%
% Or, just remove \balance and give up on balancing the last page.
%

% BALANCE COLUMNS
\balance{}

% REFERENCES FORMAT
% References must be the same font size as other body text.
\bibliographystyle{SIGCHI-Reference-Format}
\bibliography{sample}

\end{document}